# Sixteen New Linear Codes With Plotkin Sum

Fernando Hernando and Diego Ruano

*Abstract*—Sixteen new linear codes are presented: three of them improve the lower bounds on the minimum distance for a linear code and the rest are an explicit construction of unknown codes attaining the lower bounds on the minimum distance. They are constructed using the Plotkin sum of two linear codes, also called $(u|u+v)$ construction. The computations have been achieved using an exhaustiv search.

## I. INTRODUCTION

PLOTKIN sum, also called $(u|u+v)$ construction and bar product, is a classic tool to construct codes from codes already known. It was introduced in 1960 by M. Plotkin [1] and then rediscovered in [2]. However, we show that this construction can still be used to obtain codes that improve the linear codes' bounds. We have considered the tables of linear codes listed in [3]. We obtain three codes over $\mathbb{F}_4$, that improve the lower bounds of the minimum distance and thirteen codes, over $\mathbb{F}_3$ and $\mathbb{F}_4$, whose existence was known but whose construction was unknown. We also show that Plotkin bound can be used to obtain a significant amount of codes listed in [3], sometimes in a simpler way.

## II. PLOTKIN SUM

Let $C_1, C_2$ be two linear codes in $\mathbb{F}_q^n$ with parameters $[n, k_1, d_1]$ and $[n, k_2, d_2]$ respectively. The Plotkin sum of $C_1$ and $C_2$ is

$$C = \{(u, u+v) \mid u \in C_1, v \in C_2\} \subset \mathbb{F}_q^{2n}$$

One can see in [4] that $C$ is a linear code with parameters $[2n, k_1 + k_2, \min\{2d_1, d_2\}]$.

## III. NEW CODES

In [3] one can find a list with the bounds for the minimum distance of linear codes over $\mathbb{F}_q$, with $q = 2, 3, 4, 5, 7, 8, 9$, with length and dimension lower than or equal to $256, 243, 256, 130, 100, 130, 130$, respectively.

We have considered the minimum distance of the Plotkin sums of two codes whose length $n$ is in the first half of the table (we cannot compare the sum of codes in the second half), for instance for $q = 4$, $n = 1, \ldots, 121$. We have compared their minimum distance with the codes of length $2n$ in [3]. This computation can be easily and fast achieved with a simple computer program. We obtained the following new codes:

Three codes over $\mathbb{F}_4$ which improve the lower bounds:

- The plotkin sum of the codes with parameters [63,53,6] and [63,42,12] in [3] gives a [126,95,12] code, the lower bound was 11.
- The plotkin sum of the codes with parameters [64,54,6] and [64,43,12] in [3] gives a [128,97,12] code, the lower bound was 11. Furthermore, considering a shortening [4], one obtains a [127,96,≥12] code, the lower bound was also 11.

We have obtained three codes over $\mathbb{F}_3$ that give an explicit construction, which was unknown, for the lower bound:

- The plotkin sum of the codes with parameters [62,46,8] and [62,32,16] in [3] gives a [124,78,16] code. And considering a shortening, one obtains a [123,77,≥16] code.
- The plotkin sum of the codes with parameters [63,47,8] and [63,32,17] in [3] gives a [126,79,16] code.

We have obtained ten codes over $\mathbb{F}_4$ that give an explicit construction, unknown so far, for the lower bound:

- The plotkin sum of the codes with parameters [52,38,8] and [52,25,16] in [3] gives a [104,63,16] code. And, considering a shortening, one obtains a [103,62,≥16] code.
- The plotkin sum of the codes with parameters [53,39,8] and [53,26,16] in [3] gives a [106,65,16] code. And, considering a shortening, one obtains a [105,64,≥16] code.
- The plotkin sum of the codes with parameters [54,40,8] and [54,27,16] in [3] gives a [108,67,16] code. And, considering a shortening, one obtains a [107,66,≥16] code.
- The plotkin sum of the codes with parameters [61,51,6] and [61,40,12] in [3] gives a [122,91,12] code.
- The plotkin sum of the codes with parameters [62,52,6] and [62,41,12] in [3] gives a [124,93,12] code. And, considering a shortening, one obtains a [123,92,≥12] code.
- Considering a shortening of the [126,95,12] code above, one obtains a [125,94,≥12]

One can easily obtain the codes with the computer algebra system Magma [5]. For instance we construct the code [122,91,12] over $\mathbb{F}_4$, by considering the Plotkin sum of the corresponding two codes listed in [3]:

```
> F:=GF(4);
> P<x>:=PolynomialRing(F);
> a:=F.1;
```

The work of F. Hernando is supported in part by the Claude Shannon Institute, Science Foundation Ireland Grant 06/MI/006 (Ireland) and MEC MTM2007-64704 and by Junta de CyL VA025A07 (Spain). The work of D. Ruano is supported in part by DTU, H.C. Oersted post doc. grant (Denmark) and MEC MTM2007-64704 and by Junta de CyL VA065A07 (Spain)
F. Hernando is with the Department of Mathematics, University College Cork, Ireland, e-mail: F.Hernando@ucc.ie
D. Ruano is with the Department of Mathematics, Technical University of Denmark, Matematiktorvet, Building 303, DK-2800, Lyngby, Denmark, e-mail: D.Ruano@mat.dtu.dk

```
> TMP1:=BCHCode(F, 63, 5);
> TMP2:=ExtendCode(TMP1,1);
> C1:=ShortenCode(TMP2,{ 62 .. 64 });
> TMP3:=CyclicCode(65,x^21+a*x^20+a*x^19+
a*x^18+a^2*x^15+a^2*x^14+a^2*x^12+x^11+
x^10+a^2*x^9+a^2*x^7+a^2*x^6+a*x^3+a*x^2+
a*x+1);
> C2:=ShortenCode(TMP3,{ 62 .. 65 });
> C:=PlotkinSum(C1,C2);
```

Finally, we remark that a significant amount of codes in [3] can be obtained using the Plotkin sum. We compare the bounds for the minimum distance of linear codes with even length listed in [3] and show how many of them can be obtained using this sum. Sometimes, this construction is simpler than the one in [3] (considering several shortenings, puncturings, or parity check bits, ...).

| $q$ | # in [3], $n$ even | # Plotkin Sum | % |
|---|---|---|---|
| 2 | 16512 | 2676 | 16.20 |
| 3 | 14762 | 1681 | 11.38 |
| 4 | 16512 | 1350 | 8.17 |
| 5 | 4290 | 495 | 11.53 |
| 7 | 2550 | 354 | 13.88 |
| 8 | 4290 | 454 | 10.58 |
| 9 | 4290 | 431 | 10.04 |

## IV. CONCLUSION

Nine new linear codes are presented. They are obtained using the Plotkin sum, therefore we suggest that whenever a new linear code of length $n$ is obtained, one should check if it is possible to obtain a code of length $2n$, which improves the lower distance bounds, by using the Plotkin sum of that code and the codes of length $n$ in [3]. Furthermore, this computation can be easily achieved.

## ACKNOWLEDGMENT

The authors would like to thank M. Greferath for his course at Claude Shannon Institute and T. Høholdt for helpful comments on this paper.